\documentclass[twocolumn,preprintnumbers,amsmath,amssymb,10pt,a4paper,showkeys]{revtex4}
\usepackage{graphicx} \usepackage{bm}
\usepackage{amsmath,amsfonts,amssymb}

\usepackage{geometry}
\usepackage{dcolumn}

\newcommand{\ignore}[1]{}

\begin{document} 

\title{Dynamical phase coexistence:\\ a simple solution to the
  ``savanna problem''} \author{Federico Vazquez} 
\email{federico@ifisc.uib-csic.es}
\author{Crist\'obal
  L\'opez} 
\email{clopez@ifisc.uib-csic.es}
\affiliation{IFISC, Instituto de F{\'\i}sica
  Interdisciplinar y Sistemas Complejos, (CSIC-UIB), E-07122 Palma de
  Mallorca, Spain.}  
\author{Justin M. Calabrese}
\email{justin.calabrese@ufz.de}
\affiliation{Dept. of Ecological Modeling, Helmholtz Centre for
  Environmental Research-UFZ, Permoserstrasse 15, 04318 Leipzig, Germany}
\author{Miguel A. Mu\~noz} 
\email{mamunoz@onsager.ugr.es}
\affiliation{
Instituto de F{\'\i}sica Te{\'o}rica y Computacional Carlos I,
Facultad de Ciencias, Universidad de Granada, 18071 Granada, Spain}

\date{\today}

 \begin{abstract}
   We introduce the concept of {\it dynamical phase coexistence} to
   provide a simple solution for a long-standing problem in
   theoretical ecology, the so-called ``savanna problem''.  The
   challenge is to understand why in savanna ecosystems trees and
   grasses coexist in a robust way with large spatio-temporal
   variability.  We propose a simple model, a variant of the Contact
   Process (CP), which includes two key extra features: varying
   external (environmental/rainfall) conditions and tree age.  The
   system fluctuates locally between a woodland and a grassland phase,
   corresponding to the active and absorbing phases of the underlying
   pure contact process. This leads to a highly variable stable phase
   characterized by patches of the woodland and grassland phases
   coexisting dynamically. We show that the mean time to tree
   extinction under this model increases as a power-law of system size
   and can be of the order of 10,000,000 years in even moderately
   sized savannas. Finally, we demonstrate that while local
   interactions among trees may influence tree spatial distribution
   and the order of the transition between woodland and grassland
   phases, they do not affect dynamical coexistence. We expect
   dynamical coexistence to be relevant in other contexts in physics,
   biology or the social sciences.
 \end{abstract}


 \maketitle

\section{Introduction}

Savannas are open systems that feature a continuous grass layer and a
discontinuous tree layer.  They appear across a wide range of climatic
and ecological conditions, and are characterized by the stable, though
variable, coexistence of two distinct types of vegetation, trees and
grasses \cite{Sarmiento}. This coexistence is dynamic in the sense
that the density of trees varies widely both in space and time, as
recently confirmed by observations of cyclic transitions between empty
and dense tree distributions \cite{Moustakas08, Wiegand}.
  Savannas have been studied from both experimental and
theoretical points of view, and have become an important subject of
study in ecology.  However, the origin, nature, and dynamics of
savannas are not yet well understood.  How is long-term coexistence of
trees and grasses possible without the superior competitor taking
over, as happens in other ecosystems (grasslands or woodlands)?  This
is one of the long standing puzzles in theoretical ecology, commonly
referred as the {\it savanna problem}.  A tentative answer to this
question is provided by niche models in which, assuming (soil, rain,
etc) heterogeneities, each life form occupies the regions for which it
is a superior competitor \cite{Walter, Walker81, Walker82}.
 This type of solution is conceptually unsatisfactory
and is, anyhow, not supported by recent empirical observations
\cite{Jeltsch,Higgins, Bond, Scholes}.

Demographic bottleneck models  \cite{Sankaran04} invoke stochastic
explanations that rely on demographic and environmental fluctuations
to generate dynamical heterogeneities \cite{Chesson,Dodorico,Jeltsch,
  Jeltsch96, Higgins,vanWijk, Meyer}.  For example, the {\it storage
  effect} hypothesizes that birth rate variability promotes species
coexistence in communities of long-lived organisms, so that in an
environment which is frequently adverse, a long life span buffers
trees against extinction \cite{Chesson}.  Other, similar in spirit,
{\it buffering mechanisms} have also been proposed
\cite{Jeltsch}. Following these studies, we build a minimalistic
savanna model that allows us to assess the contributions of the
following features to long-term tree-grass coexistence: {\bf i)}
Variable weather conditions \cite{Jeltsch96, Higgins, vanWijk}, {\bf
  ii)} Mean annual precipitation, which has been reported to enhance
and limit the maximum tree-cover \cite{Sankaran05, Sankaran08,
  Bucini}, {\bf iii)} Adult tree longevity \cite{Higgins}, and {\bf
  iv)} Positive and negative local density-dependent tree interactions
\cite{Jeltsch96, Meyer, Mallorca}.  Some existing demographic savanna
models, which include these elements along with many others, can
reproduce the main traits of real savannas \cite{Jeltsch, Jeltsch96,
  Higgins,Sankaran04, vanWijk, Meyer}.  However, these models do not
clarify which ingredients are necessary to produce long-term
coexistence and which are superfluous.

Our goal is to construct a minimal stochastic model that explains
phase coexistence in savannas.  For that, we start from a cellular
automata defined on a square grid, in which each site can be occupied
either by one tree or grass, and that follows the dynamical rules of
the standard contact process (occupied sites are trees and empty ones
are grasses; see next section).  Two phases characterize the system:
grassland (grass only) and woodland (tree dominated).  Dynamical phase
coexistence appears when the system is driven by a {\it random
  external driver} (fluctuations in rainfall) which forces the system
to visit the two phases randomly in time. In this way, coexistence
appears in a broad region of parameter space without fine-tuning of
parameters. We shall show that the range of dynamical coexistence is
much enhanced if the {\it age of the trees} is included, so that trees
can typically endure harsh conditions.

This dynamical coexistence is not indefinite as the rainfall
fluctuations will eventually lead the system to the grassland or {\it
  absorbing state}, which is characterized by the complete absence of
trees. Unlike previous work on environmental fluctuations and
tree-grass coexistence \cite{vanWijk, Jeltsch96, Higgins}, we focus
here on characterizing the timescales over which dynamical coexistence
is robust. Our approach consists of studying stability by means of the
{\it mean life-time}, that is, the mean time that takes the savanna to
reach the final absorbing state.  We show that this time diverges with
system size as $N^\alpha$ and can be enormous for even moderately
sized systems \cite{Leigh}.

Finally, let us remark that the generic phase coexistence observed in
our model, i.e. coexistence occurring in a broad region of parameter
space, is rarely find in other non-equilibrium model/systems and, in
most cases, the domain of coexistence is typically small (see
\cite{us} and references therein).  Besides, given that the phenomenon
of coexistence
also happens in biological, physical and social sciences, the
mechanism proposed in this paper is expected to be useful in many
other contexts.

\section{Model}

Consider the simple {\it contact process} (CP) \cite{CP,AS}. Each node
$(i,j)$ of a two-dimensional square lattice can be either occupied
$z_{i,j}=1$ (tree) or vacant $z_{i,j}=0$ (grass). The dynamics is as
follows: a tree is randomly selected, and it is removed from the
system with probability $d$, otherwise, with probability $b$, it
generates an offspring, which is placed at a randomly chosen nearest
neighbor (n.n) provided it was empty (i.e. short-range
seed-dispersal). Every time a tree is selected, time $t$ is increased
by $1/N(t)$, where $N(t)$ is the total number of trees in the system;
$t$ is increased in one unit, corresponding to one Monte Carlo (MC)
step or ``year'', whenever all trees have been selected once on
average.  Fixing $d$, a phase transition appears at some critical
value $b_c$. 

For $b>b_c$ the system is in the active (the system has a
non-vanishing density of trees and can dynamically evolve) phase --
woodland --, while for $b<b_c$ it is in the absorbing -- grassland --
phase with zero tree-density (as schematically illustrated in
Fig. 1). The reason for this last name is that if the system reaches
this absorbing (i.e.  only-grass) state, the situation is
irreversible, it remains indefinitely trapped in it. Since there is no
spontaneous generation of trees, the dynamics ceases if there are only
grasses.  This type of transition is commonly termed an absorbing
phase transition \cite{AS}.

\begin{figure}
  \includegraphics[clip=true,width=70mm]{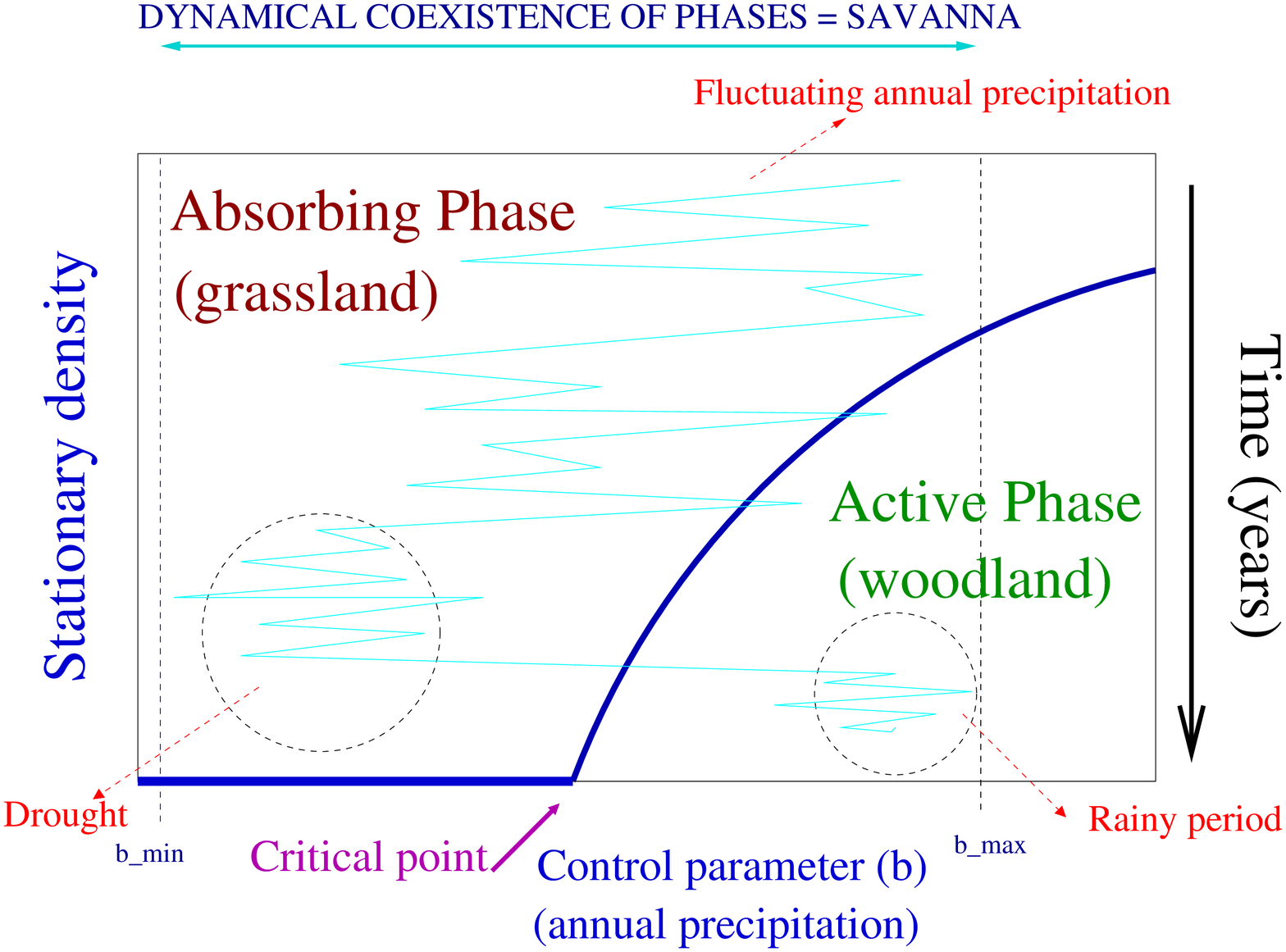}
  \caption{Schematic representation of the system dynamics. The
    underlying pure model has two homogeneous phases: active
    (woodland) and absorbing (grassland), separated by a
    critical point (the transition could also be discontinuous
    as in Fig.~\ref{rho-b}). The control parameter fluctuates in
    time and shifts from one phase to another (droughts and
    rainy periods). This, combined with long-living trees,
    prevents the system from reaching a homogeneous steady
    state.}
  \label{sketch}
\end{figure}

Now, we introduce the following two extra ingredients to model
savannas: 

{\bf i) Fluctuating external conditions}: We assume that the
birth probability depends on external conditions (mostly annual
precipitation, but also fires, etc \cite{Sankaran05}), so $b$ becomes
a time-dependent random variable (see the zigzaging line in Figure 1).
To account for possible temporal correlations in weather conditions we
take a time-correlated (colored) noise as follows.  With probability
$q$ a new value of $b$ is extracted from an uniform random
distribution in $[b_{min},b_{max}]$ at each MC step, otherwise (with
probability $1-q$) $b$ is kept fixed. Here, we take $b_{min}=0$, while
$b_{max}$ is the control parameter.  This models weather cycles of
typical length $1/q$. In a similar way, one could include more
periodic weather oscillations, as those induced by ``El Ni\~no'',
which lead to similar results.

{\bf ii) Age}: Field studies reveal that the mortality distribution of
some savanna trees is consistent across years and climatic regimes
\cite{Moustakas2006}, and that they have long lifespans. To model this
we define trees with an intrinsic age-variable, $a(i,j)$ measured in
years.  The death probability is taken to be age-dependent:
$d\rightarrow d(a(i,j))$.  In particular, a random number, $\eta$, is
extracted from a Gaussian distribution of mean $a_m$ and variance
$\sigma$ (typically, $a_m=100$ and $\sigma=20$).  If $a(i,j) \ge \eta$
then the selected old tree is removed; otherwise nothing happens.

Some other effects, such as density-dependent demographic rates, can
also be easily implemented in the model.

{\bf Density-dependence:} Negative and positive local
density-dependent death probabilities account for tree-tree {\it
  competition} and {\it facilitation}, respectively.  Both of these
effects have been reported to act in savannas \cite{Mallorca}. To
model competition between a {\it young} tree at site $(i,j)$ (tree
with age  $a(i,j) \leq a_{est}$, where $a_{est}$ is the {\it
  establishment age}) and its neighbors, we increase its death rate as
a function of the number of its nearest neighbors trees
$NN(i,j)=\sum_{(k,l)}z_{k,l}$, where the sum is restricted to nearest
neighbors sites of $(i,j)$.  Thus, we take $d(i,j) = 1 -
\exp\left[-NN(i,j)\right]$.  Contrarily, to model strong facilitation
we consider $d(i,j)=1$ for $NN(i,j)=0$ and  $1$, and $d(i,j)=0$ for
$NN(i,j)\geq 2$, that is, trees born in sites with only a few occupied
neighbors die with certainty, otherwise they survive.

\section{Model analysis and Results}

We first analyze the role of each ingredient we have added to the
basic contact process separately.  Taking a fluctuating $b$ and a
fixed death rate, i.e. {\bf switching off the age effect}, the system
shifts randomly between the tendencies to be in the active (tree
density larger than zero) and in the absorbing (zero tree density)
phase of the underlying pure model (notice the zigzagging trajectory
in Fig.~\ref{sketch}).  For $b_{max} >b_c$, the system hovers around
its critical point, while, if $b_{max} < b_c$ (resp.  $b_{min} >b_c $)
the fluctuating system is in the absorbing (active) phase, i.e. it is
a grassland (woodland). In principle, if the time series of $b$
happens to be adverse (i.e.  $b<b_c$) for a sufficiently long time
interval, any finite system falls into the absorbing state; i.e. {\it
the system has variability but little resilience to long adverse periods}.

\begin{figure}
  \includegraphics[clip=true,width=70mm]{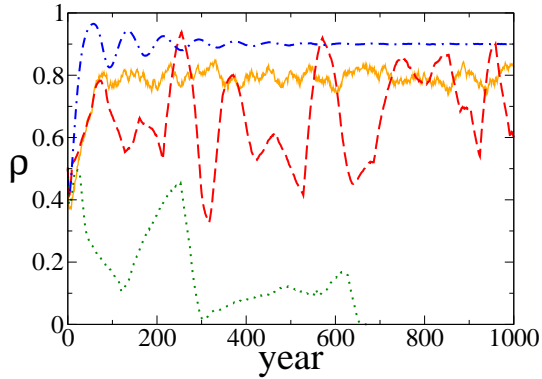}
  \caption{Time evolution of the tree-density in different cases.  All
    curves except the dotted one ($a_m=30$, $\sigma=6$) are for
    $a_m=100$, $\sigma=20$.  The dashed-dotted curve ($q=0$, i.e.
    model without variability) exhibits damped oscillations. The solid
    line ($q=1$) shows small variability, while the dashed and dotted
    curves are for $q=0.03$ (intermediate variability).  Notice that
    the one with smaller maximum age ($a_m=30$; dotted curve) does not
    survive to an adverse period, while its analogous for large age
    ($a_m=100$; dashed curve) does.}
\label{rho-t}
\end{figure}

The effect of weather correlations is as follows: for $0 < q <<1 $ the
birth rate is constant for long periods and the system typically
jumps, every $1/q$ years, from a pure CP homogeneous state to another
one and, therefore, when $b$ takes a value smaller than $b_c$ it falls
ineluctably into the absorbing state.  For $q \approx 1$, the rate of
variation of $b$ is very fast and the system does not have the time
required to relax to any pure CP steady state, and reaches instead an
averaged density value (see solid curve in Fig.~\ref{rho-t}). For
realistic intermediate values (e.g. $q=0.03$), the system exhibits
much larger oscillations (see the dashed and the dotted line in
Fig.~\ref{rho-t}) which resemble those in real savannas
\cite{Moustakas08}.

On the other hand, taking a fixed birth rate, $q=0$, i.e. {\bf
  switching off the weather variability} the model becomes a CP with
age. For this, if trees die at a fixed given maximum age, $a_m$ (i.e.
with $\sigma=0$), the density is known to exhibit damped oscillations
in time of period $2 a_m$, and to converge asymptotically to a
homogeneous stationary value \cite{Iran} (see dashed-dotted line of
Fig.~\ref{rho-t}).  Our model (with $\sigma >0$) exhibits analogous,
though more variable, damped oscillations, and converges either to the
absorbing state or to an active homogeneous state.  Separating these
two regimes there is a phase transition.  In this case {\it the system
  is resilient but has little variability}.  Obviously resilience
grows with the maximum age.

As we demonstrate below, the full model, which includes both age and
correlated variable rainfall, exhibits variability and resilience (see
the dashed and dotted curves in Fig.~\ref{rho-t}); the larger the
maximum age, the larger the resilience.  The system fluctuates locally
between the absorbing and the active phases, but it is able to
preserve ``islands'' of the unfavored phase in a ``sea'' of the
dominant phase, as is required for generic phase coexistence. This is
the basic mechanism of {\it dynamical phase coexistence}.

\begin{figure}
\includegraphics[clip=true,width=70mm]{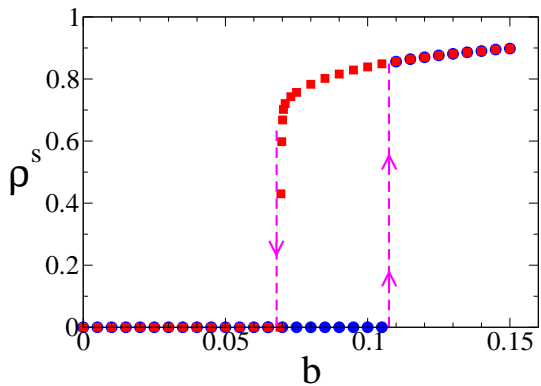}
\caption{Phase diagram for the underlying pure model ($q=0$) in the
  case of strong facilitation; observe the discontinuous transition
  and the hysteresis loop.}
\label{rho-a}
\end{figure}

\begin{figure}
\includegraphics[clip=true,width=70mm]{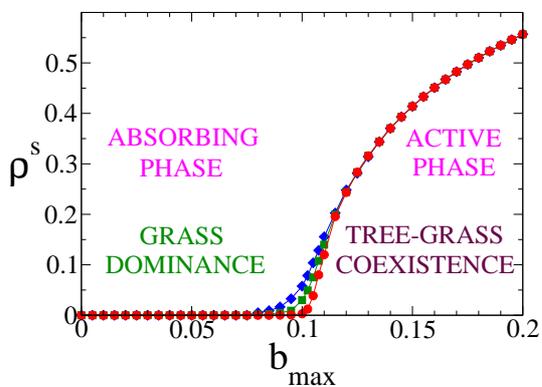}
\caption{Stationary tree-density $\rho^s$ vs maximum birth probability
  $b_{max}$, averaged over surviving realizations for the model with
  competition, $q=0.1$, $a_m=100$, $a_{est}=1$, and system sizes
  $N=40^2$ (diamonds), $N=80^2$ (squares) and $N=320^2$ (circles).}
\label{rho-b}
\end{figure} 

The effect of {\bf density-dependence} is as follows. Computer
simulations show that, the underlying pure model ($q=0$ and no age),
either in the absence of density dependence and in the case of
competition or weak facilitation, exhibits a continuous absorbing
phase transition (as schematized in Fig.~\ref{sketch}). Instead, for
strong facilitation the underlying transition can be discontinuous
with a broad hysteresis loop (see Fig.~\ref{rho-a}), implying that
around the transition the two dynamically coexisting phases are very
different. 

Fig.~\ref{rho-b} shows the stationary tree density as a function
$b_{max}$ ($q=0.1$, $a_m=100$, $a_{est}=1$) illustrating the existence
of an active phase and a continuous phase transition in the full model
with competition.  The active phase of the full model is the phase of
coexistence: the two phases of the pure model (grass and wood)
coexist.  A very similar continuous transition is obtained in the case
of strong facilitation; even if the underlying pure-model transition is
discontinuous, once varying conditions are switched on, the transition
between the absorbing and the tree-grass coexistence phases becomes
continuous.

\begin{figure}
  \includegraphics[clip=true, width=70mm]{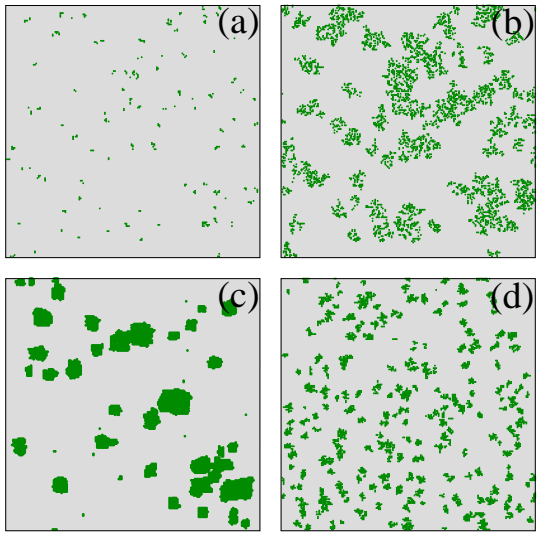}
  \caption{Snapshots of the system for different versions of the model
    with $q=0.01$ and $b_{max}=1$.  (a) and (b) correspond to the same
    realization with $a_{est}=4$, competition, and two different
    times.  The system fluctuates locally and globally between small
    and large densities, as in real savannas.  (c) Facilitation,
    $a_{est}=4$. (d) Competition, $a_{est}=1$.}
\label{snapshots}
\end{figure}

For illustration purposes, Fig.~\ref{snapshots} shows snapshots of
such a phase for different parameters at different times. Panels (a)
and (b) correspond to the case of competition, and both have the same
parameter values but look quite different, illustrating the large
spatio-temporal variability.  Panel (c) shows that, in the case of
facilitation more compact clusters are observed (as justified by the
underlying discontinuous transition).  These snapshots are visually
very similar to pictures of real savannas \cite{Moustakas08}; notice the
presence of irregularly distributed tree clusters of different sizes
and shapes. Panel (d) corresponds to competition, but with slightly
different parameters. We can conclude that density-dependence controls
the shape of emerging clusters, but it is not an essential ingredient
for dynamical coexistence.

\begin{figure}
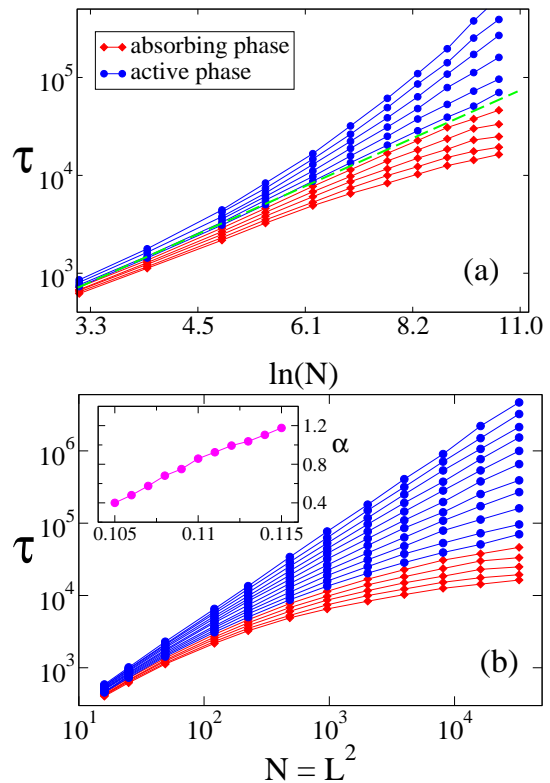

\includegraphics[clip=true,width=70mm]{Fig-6.eps}
\includegraphics[clip=true,width=70mm]{Fig-7.eps}
\caption{(a) Savanna mean life-time $\tau$ vs the logarithm of
  its size $N$ on a log-log scale, for
  different values of $b_{max}$ below (diamonds, $b_{max} =
  0.100-0.104$) and above (circles, $b_{max} = 0.105-0.115$)
  the transition point $b_{max}^c \simeq 0.1045$. At $b_{max}^c$ $\tau$
  grows as $(\ln N)^{3.68}$  (straight dashed line). (b) $\tau$ vs $N$
  on a log-log scale.  $\tau$ diverges as $N^{\alpha}$ for $b_{max} >
  b_{max}^c$ (circles), with the exponent $\alpha$ proportional to
  $b_{max}$ (inset), indicating that the active phase is stable when
  $N \to \infty$.}
\label{tau}
\end{figure} 

Let us now scrutinize the stability of the active phase, where there
is dynamical coexistence: how long does it persist through periods of
adverse rainfall conditions?

In a single realization, independent of the value of $b_{max}$ and provided
that $b_{min}$ is smaller than the critical values of the pure model, the
density of trees reaches the absorbing state $\rho=0$, that consists of a
totally empty lattice where no more trees are born, and therefore the
evolution stops.  In other words, trees go extinct in a time that depends on
the specific realization, due to stochastic fluctuations, and the size of the
system.  We have computed the \emph{mean life-time}, $\tau$, that is, the
average over many realizations of the time required to reach tree extinction,
for several values of $b_{max}$ and different system sizes $N$ (see
Fig.~\ref{tau}).  The value $b_{max}^c \simeq 0.1045$ at which $\tau$ grows
as $\tau \sim (\ln N)^{3.68}$ [straight dashed line in Fig.~\ref{tau}(a)],
agrees within error-bars with the point where the stationary density $\rho^s$
goes to zero in Fig.~\ref{rho-b}, thus we take it as the active-absorbing
phase transition point.  In Fig.~\ref{tau}(b) we observe that, remarkably, in
the active phase (circles) $\tau$ diverges with $N$ as a power law $N^\alpha$
(with possible log corrections), where the exponent $\alpha$ increases with
$b_{max}$ (see inset). That is, the mean life-time increases very rapidly with
system size, and therefore one expects that real (large) savannas survive for
very long periods.  

As an illustration, consider a $1000$ hectare savanna,
which corresponds to a $632 \times 632$ square lattice (assuming, say,
neighboring lattice sites are separated by $5$ meter distance), taking
$b_{max} \simeq 0.111$ not very deep into the active phase, then our model
predicts that its expected time to extinction is of the order of $10$ million
years.  This leads to the conclusion that the coexistence of grassland and
woodland phases is stable for extremely long time periods. Furthermore, we
find that the exponent $\alpha$ increases with the maximum tree age $a_{m}$
(not shown), so that age strongly stabilizes the active phase and increases
the average savanna mean life-time time.

\section{Empirical example}

\begin{figure}
  \includegraphics[clip=true,width=70mm]{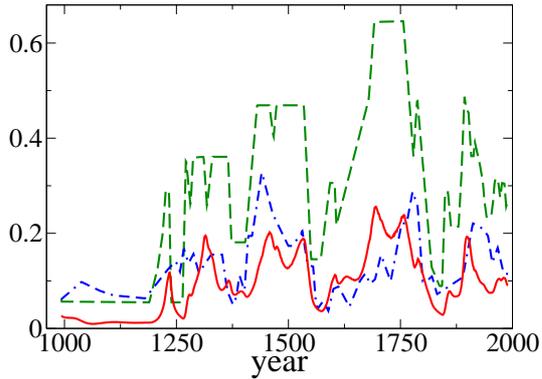}
  \caption{Ratio of tree to grass pollen in the Crescent Island Crater
    core NC93 in Kenya (dashed-dotted line) and from numerical
    simulations of the model (solid line), on a $100 \times 100$
    square lattice, with age $=100$.  The time series of the birth
    probability in the simulations (dashed line) was taken to be
    proportional to the depth of the lake Naivasha, located next to
    the Island Crater, in Kenya.}
\label{ratio}
\end{figure}

To demonstrate that this simple model is a reasonable descriptor of
savanna dynamics, we fit it to a paleoecological time series of
tree-grass pollen ratios, $R_{data}(t)$,  and lake depth levels,
$\delta(t)$, from the Crescent Island Crater site in central Kenya
[Lamb et al. 2003, Vershuren et al. 2000]. The long-term nature of
this time series ($>1000$ years) allows us to examine a much wider
range of density fluctuations than is possible with short-term,
directly observed tree density data. The cost is that these
paleoecological data sets rely on proxy measures (e.g. fossil pollen
for vegetation, fossil midge and diatom assemblages for lake depth)
and are thus necessarily much less precise than direct measures of
tree density and rainfall. We digitized figures 3B and 3C in [Lamb et
  al. 2003] to obtain the pollen ratio and lake depth time series,
respectively, using the program Engauge V 4.1
[http://digitizer.sourceforge.net/]. As these time series were both
reconstructed at the same site, over the same time interval, they give
us rough proxy measures for both precipitation patterns and tree
density at this site going back over 1000 years (see
Fig.~\ref{ratio}).

We ran simulations of the model with a birth probability given by the
expression $b(t) = \gamma\,\delta(t)$ (where $\gamma$ is a constant),
that is, assuming that the amount of rainfall, and therefore $b(t)$,
is proportional to the lake's depth.  We included neither age nor
density effects.  We also considered that the amounts of pollen from
trees and grasses are proportional to the number of trees $\rho L^2$
(with $L=100$) and the area covered by grass $(1-\rho) L^2$,
respectively, thus we took $R_{num}(t)=\beta\,\rho/(1-\rho)$ as the
tree-grass pollen ratio in the simulations, where $\beta$ is a
constant.  Then, for given values of $\gamma$ and $\beta$, we averaged
over $10$ independent realizations of the dynamics, and calculated the sum of
squared deviations between the model and the tree-grass ratio data
over the $1000$ years period as 
$\sum_{i=1}^{1000}\left[R^i_{data}(t)-R^i_{num}(t)\right]^2$. 

Comparing the model to the paleoecological data from the Crescent
Island Crater site, we found by numerical direct search that
$\gamma_{min} = 0.018$ and $\beta_{min} = 0.005$ are the values that
minimized the sum of squared errors. As can be seen in
Fig.~\ref{ratio}, the time evolution of the tree-grass pollen ratio
from numerical simulations (solid line) is strongly biased by the
temporal function $b(t)$ (dashed line), and it has only a qualitative
agreement with the evolution of the real ratio (dashed-dotted line).
We have checked that the age of trees in the model does not affect the
evolution of the ratio very much, and that single realizations are
typically very similar, given that fluctuations coming from the system
size are much smaller than fluctuations induced by the variation of
$b(t)$.  We also note in Fig.~\ref{ratio} that the real-data pollen
ratio also roughly follows the variation of the lake's depth. These
results demonstrate that this correlation between rainfall and pollen
ratio, already noted in \cite{Lamb,Verschuren}, is correctly described
by our very simple model. This correspondence is especially
encouraging given that we have manipulated only two parameters here to
fit the model.

\section{Summary}

Here we have shown that a simple extension of the contact process is
capable of providing a robust and general explanation for tree-grass
coexistence in savannas. Specifically, fluctuating rainfall levels
force the system to oscillate between a tree-dominated woodland
(active) phase and a grass-only (absorbing) phase; a behavior we call
dynamical phase coexistence. Dynamical coexistence is not permanent,
but, as we have shown, it is expected to last for geologically
significant periods of time. In other words, the mechanism discussed
here facilitates coexistence over periods of time easily long enough
to span the gaps between major disturbance events, such as ice ages,
that can create and destroy savannas.  

Strictly speaking, the only essential ingredient for dynamical phase
coexistence is the presence of externally varying conditions. Said
another way, dynamical coexistence is independent of the lifespans of
trees, the degree of correlations in weather conditions, and the
presence and nature of density-dependent local interactions among
trees. Adding age-dependent tree death to the model greatly enhances
the stability and robustness of coexistence.  Adding weather
correlations of intermediate magnitude increases the variability in
tree cover and can help the model mimic the wide variability in tree
cover that has been observed among sites with similar rainfall
\cite{Sankaran05, Bucini, Sankaran08}.  Finally, adding local
tree-tree interactions can allow the model to reproduce a range of
tree spatial patterns that have been observed in real savannas,
including dense thickets of trees and more widely spaced and open
configurations \cite{Mallorca, Moustakas08}.  Thus these additional
features affect the nature of several observable features of savannas
but are not the primary drivers of tree-grass coexistence.

This last point merits further discussion. While several stochastic
savanna models have included varying weather conditions and many other
factors \cite{Jeltsch96, Jeltsch98, Higgins, Meyer, vanWijk}, none
has, to our knowledge, clearly identified the minimal conditions that
facilitate tree-grass coexistence. Our strategy of starting from the
well-studied contact process and extending the model in a stepwise
fashion has allowed us to identify the contribution of each model
component to tree-grass coexistence and to other features observed in
real savannas. We can therefore unambiguously state that fluctuating
external conditions alone are sufficient to facilitate long-term
(though not indefinite) tree-grass coexistence. A further advantage of
our minimalistic approach is that it does not require parameter fine
tuning to achieve coexistence, as has been the case in other
stochastic savanna models  \cite{Jeltsch96, Higgins}.  Finally, unlike
previous studies, we have been able to clearly demonstrate the time
scales over which dynamical phase coexistence persists.

Despite the simplicity of our model, we have shown that it is capable
of describing qualitatively a 1000 year paleoecological dataset when
it is given a quantity proportional to rainfall as a driving
input. The key result here is that the model is able to reproduce the
correlation between weather patterns and tree-grass ratios. The rough
quantitative agreement between the model and data is not surprising
given that these data, which are based on proxy measures instead of
direct observation, are relatively crude and that only two parameters
of our dynamic, \emph{nonlinear} model, were tuned to achieve the
fit. We note that though empirical data are frequently used in savanna
modeling studies, such studies typically only consider single
snapshots in time, or very short time series  \cite{Jeltsch99,
  Higgins, vanWijk}.

While we have focused here on savannas, the concept of dynamical phase
coexistence is much more general than that. Variants of the Contact
Process are used in many different fields including: Physics, Ecology,
Epidemiology, Sociology, etc. Furthermore, it seems likely that many
of the systems to which these models are applied may feature some
degree of forcing by externally varying conditions. Thus, we expect
that the concept of dynamical phase coexistence will find broad
applicability in a range of scientific disciplines.

\section{Acknowledgment}
F.V., C.L, and J.M.C. acknowledge support from NEST-Complexity project
PATRES (043268). F.V. and C.L. acknowledge support from project
FISICOS (FIS2007-60327) of MEC and FEDER. M.A.M. acknowledges financial
support from the Spanish MICINN-FEDER under Project N. FIS2009-08451 and from
Junta de Andaluc{\'\i}a as group FQM-165.


\begin{thebibliography}{10}


\bibitem[Bond 2008]{Bond} Bond, W. J. 2008.
What limits trees in C4 grasslands
and savannas? 
Annual Review of  Ecology, Evolution, and Systematics
39, 641–659.

\bibitem[Bucini and Hanan 2007]{Bucini} Bucini, G., and Hanan, N.P., 2007.
A continental-scale analysis of tree cover in African savannas.
 Global
 Ecology and Biogeography 16, 593-605.

\bibitem[see Calabrese et al. In press] {Mallorca} Calabrese, J.M,
  Vazquez, F. L\'opez, C., San Miguel, M., and Grimm, V., 2009.  The
  individual and interactive effects of tree-tree establishment
  competition and fire on savanna structure and dynamics.  Am. Nat.,
  in press.

\bibitem[Chesson 2000]{Chesson} Chesson,  P. L., 2000.
 Mechanisms of maintenance of species diversity.
Annu. Rev. Ecol. Syst.  31, 343-358;
 Warner, R.,  and Chesson, P.L., 1985. 
Coexistence mediated by
recruitment fluctuations: a field guide to the storage effect.
 Am. Nat. 125, 769-787.

\bibitem[D'Odorico et al. 2007]{Dodorico} 
D'Odorico, P., Laio, F., Porporato, A., Ridolfi, L.,  and Barbier, N., (2007).
 Noise-induced vegetation patterns in fire-prone savannas, 
J. Geophys. Res. 112, G02021, doi:10.1029/2006JG000261. 

\bibitem[Gacs 2001]{Gacs} G\'acs, P., 2001.
Reliable cellular automata with self-organization.
 J. Stat. Phys. 103, 45-267.

\bibitem[Gerami 2002]{Iran} Gerami, R., 2002.
Criticality and oscillatory behavior in
a non-Markovian contact process.
 Phys. Rev. E 65, 036102-036105.

\bibitem[Harris 1974]{CP}  Harris, T.E., 1974.
Contact interactions on a lattice.
 Ann. Prob. 2, 969-988.

\bibitem[Higgins et al. 2000]{Higgins} 
Higgins, S.I., Bond W.J. and Trollope, W.S.W., 2000.
Fire, resprouting and variance: a recipe for grass-tree coexistence in savanna.
Journal of Ecology 88, 213-229

\bibitem[Hinrichsen 2000]{AS} Hinrichsen, H., 2000.
Nonequilibrium critical phenomena and phase transitions into
absorbing states.
Adv. Phys. 49, 815-958.  

\bibitem[Jeltsch et al. 1996]{Jeltsch96}  Jeltsch, F., Milton, S.J.,
  Dean, W.R.J., and VanRooyen, N.,. 1996. Tree spacing and coexistence
  in semiarid savannas. Journal of Ecology 84, 583-595.

\bibitem[Jeltsch et al. 1998]{Jeltsch98}  Jeltsch, F., Milton, S.J.,
  Dean, W.R.J., van Rooyen, N., and Moloney, K.A., 1998. Modelling the
  impact of small-scale heterogeneities on tree-grass coexistence in
  semi-arid savannas. Journal of Ecology 86, 780-793.

\bibitem[Jeltsch et al. 1999]{Jeltsch99}  Jeltsch, F., Moloney, K.,
  and Milton, S.J., 1999. Detecting process from snapshot pattern:
  lessons from tree spacing in the southern Kalahari. Oikos 85,
  451-466.

\bibitem[Jeltsch et al. 2000]{Jeltsch} Jeltsch, F.,  Weber, G.E., and
  Grimm, V., 2000.  Ecological buffering mechanisms in savannas: A
  unifying theory of long-term tree-grass coexistence.  Plant Ecology
  161, 161–171. doi: 10.1023/A:1026590806682.

\bibitem[Jensen 1996]{Jensen} Jensen, I., 1996.
Temporally disordered bond percolation on the
directed square lattice.
Phys. Rev. Lett. 77, 4988-4991.

\bibitem[Lamb et al. 2003]{Lamb} Lamb, H.,  Darbyshire I.,  and
  Verschuren, D., 2003.  Vegetation response to rainfall variation and
  human impact in central kenya during the past  1100 years.  The
  Holocene 13(2), 285-292.

\bibitem[Leigh 1981]{Leigh}  Leigh, E.G., Jr., 1981.
The average lifetime of a population in  random environment,
J. Theor. Biol. 90, 213-239.

\bibitem[Meyer et al. 2007]{Meyer} Meyer, K. M., Wiegand, K., Ward, D., and
   Moustakas, A.,  2007. SATCHMO: A spatial simulation model of growth,
  competition, and mortality in cycling savanna patches. Ecological
  Modelling 209, 377-391;  Moustakas, A., Guenther, M.,  Wiegand, K.,
 Mueller, K.H., Ward, D.,  Meyer, K.M.,  and Jeltsch, F., 2006. Long-term
  mortality patterns of the deep-rooted Acacia erioloba: the middle
  class shall die! Journal of Vegetation Science 17, 473-480.

\bibitem[Moustakas et al. 2006]{Moustakas2006} Moustakas A., Gunther
  M., Wiegand K., Muller K.-H., Ward D. and Meyer K.M., (2006).
  Mortality of Acacia erioloba: influence of climate and
  intra-specific competition.  The middle class shall die!  Journal of
  Vegetation Science, 17, 473-480.

\bibitem[Moustakas et al. 2008]{Moustakas08}  Moustakas, A.,  Wiegand,
  K.,  Getzin, S.,  Ward, S.D., Meyer, K.M.,  Guenther, M., and
  Mueller, K.H., 2008. Spacing patterns of an Acacia tree in the
  Kalahari over a 61-year period: how clumped becomes regular and vice
  versa. Acta Oecologica 33, 355-364.

\bibitem[Mu\~noz et al. 2005]{us} M. A. Mu\~noz, M.A., de los Santos, F., and
Telo da Gama, M., 2005.
Generic coexistence of stable phases in nonequilibrium systems.
Eur. Phys. Jour. 43, 73-79, DOI: 10.1140/epjb/e2005-00029-3.

\bibitem[Sankaran et al. 2004]{Sankaran04} Sankaran, M., J. Ratnam, and
  Hanan, N.P., 2004. Tree-grass coexistence in savannas revisited -
  insights from an examination of assumptions and mechanisms invoked
  in existing models. Ecology Letters 7, 480-490.

\bibitem[Sankaran et al. 2005]{Sankaran05} Sankaran, M., Hanan, N.P.,
  Scholes, R.J., Ratnam, J., Augustine, D.J., Cade, B.S., Gignoux, J.,
  Higgins, S.I., Le Roux, X., Ludwig, F.,  Ardo, J., Banyikwa, F.,
  Bronn, A., Bucini, G., Caylor, K. K. Coughenour, M.B., Diouf, A.,
  Ekaya, W., Feral, C.J., February, E.C. Frost, P.G.H.,  Hiernaux, P.,
  Hrabar, H., Metzger, K.L., Prins, H.H.T., Ringrose, S., Sea, W.,
  Tews, J., Worden, J., and Zambatis, N., 2005. Determinants of woody
  cover in African savannas. Nature 438, 846-849.

\bibitem[Sankaran et al. 2008]{Sankaran08} Sankaran, M., Ratnam, J.,
  and Hanan, N., 2008. Woody cover in African savannas: the role of
  resources, fire and herbivory. Global Ecology and Biogeography 17,
  236–245.  

\bibitem[Sarmiento 1984]{Sarmiento}  Sarmiento, G., 1984. {\it The
  ecology of Neotropical savannas}, Harvard University Press,
  Cambridge.  Scholes, R.J., and  Archer, S.R., 1997.  Tree-grass
  interactions in savannas.  Ann. Rev. Ecol. Syst. 28, 517.

\bibitem[Scholes and Walker 1993]{SW} Scholes, R. J.,  and Walker,
  B.H., 1993.  An African Savanna: synthesis of the Nylsvley study,
  Cambridge University Press, Cambridge.

\bibitem[Scholes and Archer 1997]{Scholes} Scholes, R.J., and Archer,
  S.R., 1997. Tree-grass interactions in savannas. Annual Review of
  Ecology and Systematics 28, 517-544.

\bibitem[van Wijk and Rodriguez-Iturbe 2002]{vanWijk} van Wijk, M. T.,
  and Rodriguez-Iturbe, I.,  2002.  Tree-grass competition in space
  and time: Insights from a simple cellular automata model based on
  ecohydrological dynamics.  Water Resour. Res., 38(9), 1179,
  doi:10.1029/2001WR000768. 

\bibitem[Verschuren et al. 2000]{Verschuren}  Verschuren, D.,  Laird,
  K.R.,  and Cumming, B.F., 2000.  Rainfall and drought in euqatiorial
  east Africa during the past 1110 years.  Nature 403,  410-413.

\bibitem[Walter 1971]{Walter} Walter, H., 1971.  Ecology of Tropical
  and sub-tropical vegetation, Oliver and Boyd, Edinburg.

\bibitem[Walker et al. 1981]{Walker81}  Walker, B.H., Ludwig, D.,
  Holling, C.S., and Peterman, R.M., 1981. Stability of semi-arid
  savanna grazing systems. Journal of Ecology 69, 473-498.

\bibitem[Walker and Noy-Meir 1982]{Walker82} Walker, B.H., and
  Noy-Meir, I., 1982. Aspects of the stability and resilience of
  savanna  ecosystems.  In: Walker BJ, Huntley BH (eds) Ecology of
  tropical savannas. Springer, Berlin, 556–590 .

\bibitem[Wiegand et al. 2006]{Wiegand} Wiegand, K., Saitz, D.,  and
  Ward, D., 2006. A patch-dynamics approach to savanna dynamics and
  woody plant encroachment - insights from an arid
  savanna. Perspectives in Plant Ecology Evolution and Systematics 7,
  229-242.



\end{thebibliography}
\end{document}